\documentclass{article}

\usepackage[utf8]{inputenc}
\usepackage{amsmath}
\usepackage{amsfonts}
\usepackage{amssymb}
\usepackage{amsthm}
\usepackage{graphicx}
\usepackage{caption}
\usepackage{subcaption}
\usepackage{hyperref}
\hypersetup{colorlinks = true, unicode=true,  urlcolor=blue, linkcolor=blue, filecolor=blue, citecolor=blue}    % colorlinks=true, citecolor=red
\usepackage{authblk}
\usepackage{footmisc}
\usepackage{natbib}

\title{\bf A note on matricial ways to compute Burt's structural holes in networks\thanks{email: alessio.muscillo2@unisi.it -- ORCID iD: \href{https://orcid.org/0000-0002-2648-4272}{0000-0002-2648-4272}
\\ The author acknowledges funding from the Italian Ministry of Education \emph{Progetti di Rilevante Interesse Nazionale} (PRIN) grant 2017ELHNNJ.\\ The author thanks Paolo Pin, Tiziano Razzolini, Claudia Ruzza and Gabriele Lombardi for their help and support. Additional material and Python code used in this note are available online \href{https://www.dropbox.com/s/v4qtzx5ieirzaib/Appendix\%20-\%20A\%20note\%20on\%20matricial\%20ways\%20to\%20compute\%20Burt\%27s\%20structural\%20holes\%20-\%20by\%20Alessio\%20Muscillo.pdf?dl=0}{here}.}}

\author{Alessio Muscillo}
\affil{Department of Economics and Statistics, Universit\`a di Siena, piazza San Francesco 7, 53100, Italy}

\date{\today}

\begin{document}

\maketitle

\begin{abstract}
    \noindent
    In this note I derive simple formulas based on the adjacency matrix of a network to compute measures associated with Ronald S. Burt's structural holes (effective size, redundancy, local constraint and constraint).
    This can help to interpret these measures and also to define naive algorithms for their computation based on matrix operations.\vspace{2mm}\\
    \noindent
    \textbf{Keywords}: network measures, structural holes, effective size, redundancy, constraint, computation
    % \vspace{1mm}
    
    % \noindent
    % \textbf{JEL classification codes}: C63, D85, L14 
\end{abstract}

\section{Introduction}

In the last decades, the social context in which economic activities are embedded has become more and more the focus of attention and research \citep{granovetter1985economic,schweitzer2009economic,goyal2018heterogeneity}.
Regularities of network structures that shape -- and, in turn, are shaped by -- economic behavior have been studied with the increased awareness that ``designing many economic policies requires a deep understanding of social structure'' \citep{jackson2017economic}.

% To analyze how the network structure relates to economic behavior, great attention is devoted to studying measures able to capture an individual's importance, influence, or centrality in a network. 
To analyze how the network structure relates to economic behavior, different notions and measures are used to capture an individual's importance, influence, or centrality in a network \citep{bloch2019centrality}. 
% One of the most fascinating notions is that of \emph{structural holes}, developed by \citet{burt2009structural}, which refers to the absence of connections between groups and has been linked to the fact that filling voids and bridging gaps might be beneficial for individuals able to do so.
One of the most fascinating concept is that of \emph{structural holes}, developed by \citet{burt2009structural}, which refers to the absence of connections between groups and to the fact individuals might benefit from establishing links that fill voids and bridge gaps.
The versatility of this concept has stimulated the definition of several measures, each capturing different aspects in different applications \citep{burt2004structural,goyal2007structural,rubi2017structural}. 
However, this has also generated confusion when it comes to which exact measure has to be computed, how to compute it and what are the relations with similar measures \citep{borgatti1997structural,newman2018networks,everett2020unpacking}.
Moreover, calculating these measures directly by applying the definition formulas can be very slow and computationally intensive, because it would require looping over each node's neighbors (and its neighbors' neighbors). 

\begin{figure}[tb]
    \centering
    \caption*{\bf Comparison with NetworkX's routines}
	\begin{subfigure}[t]{.49\textwidth}
		\includegraphics[width=\textwidth]{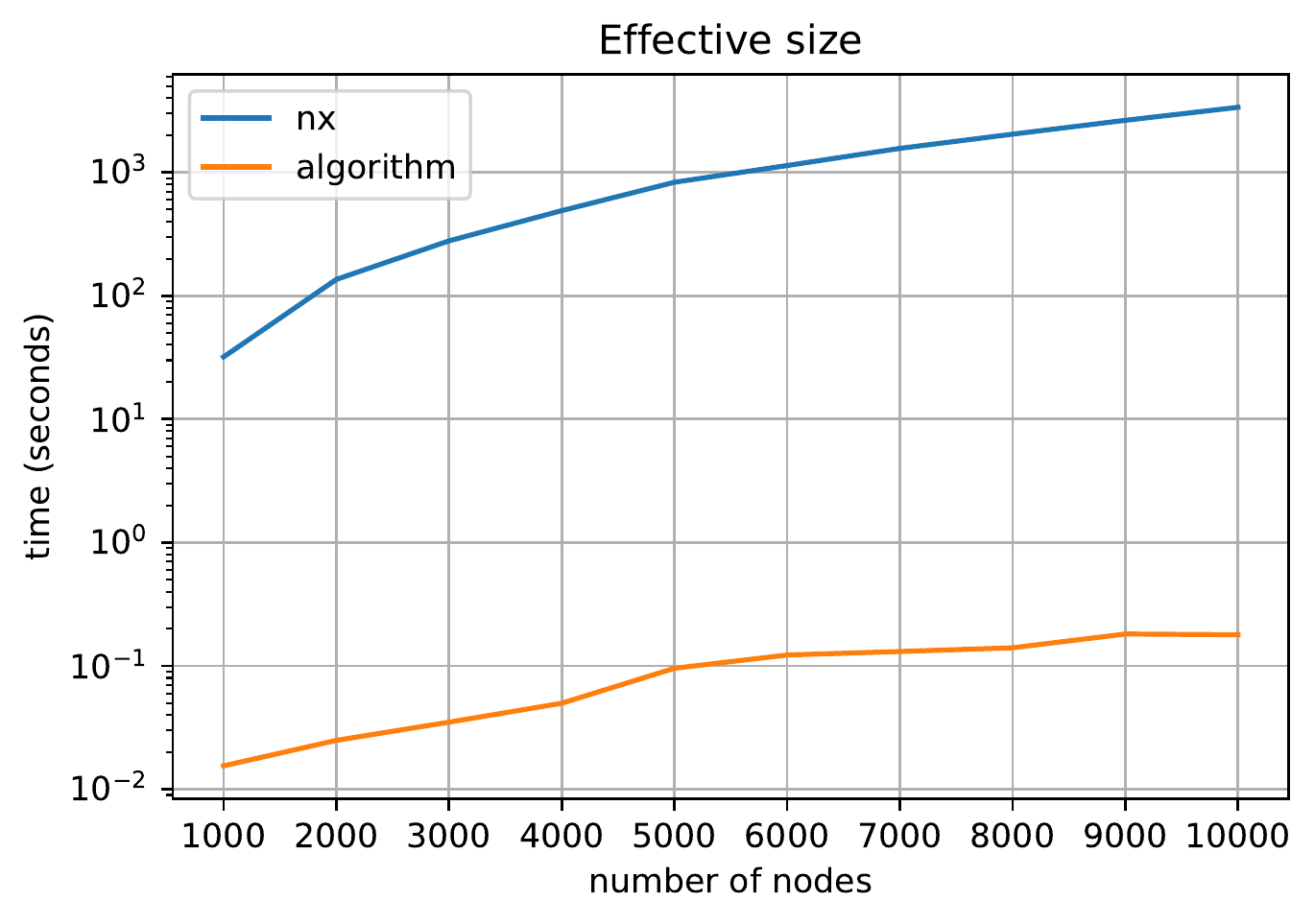}
	\end{subfigure}
	\begin{subfigure}[t]{.49\textwidth}
		\includegraphics[width=\textwidth]{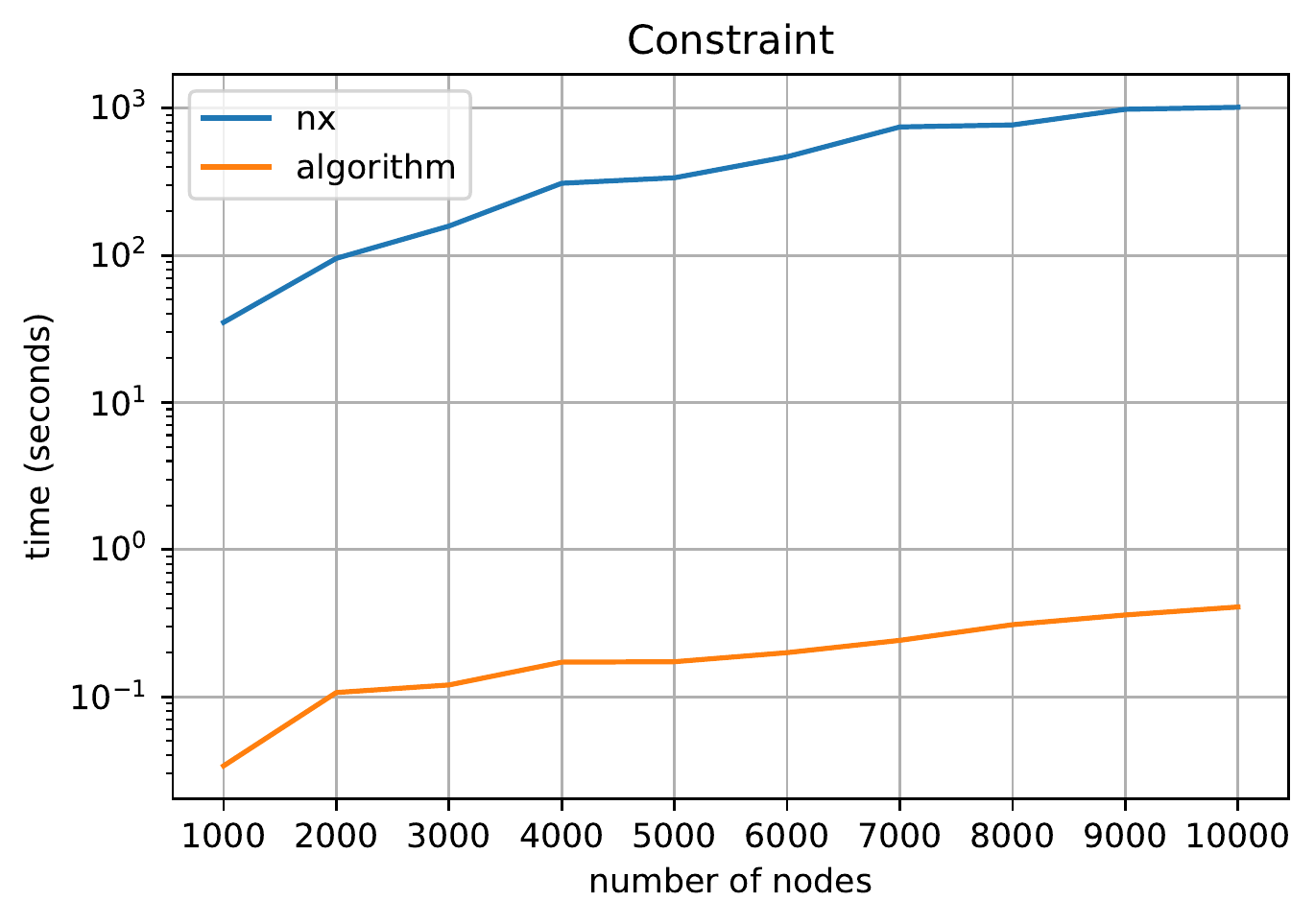}
	\end{subfigure}
	\begin{subfigure}[t]{.49\textwidth}
		\includegraphics[width=\textwidth]{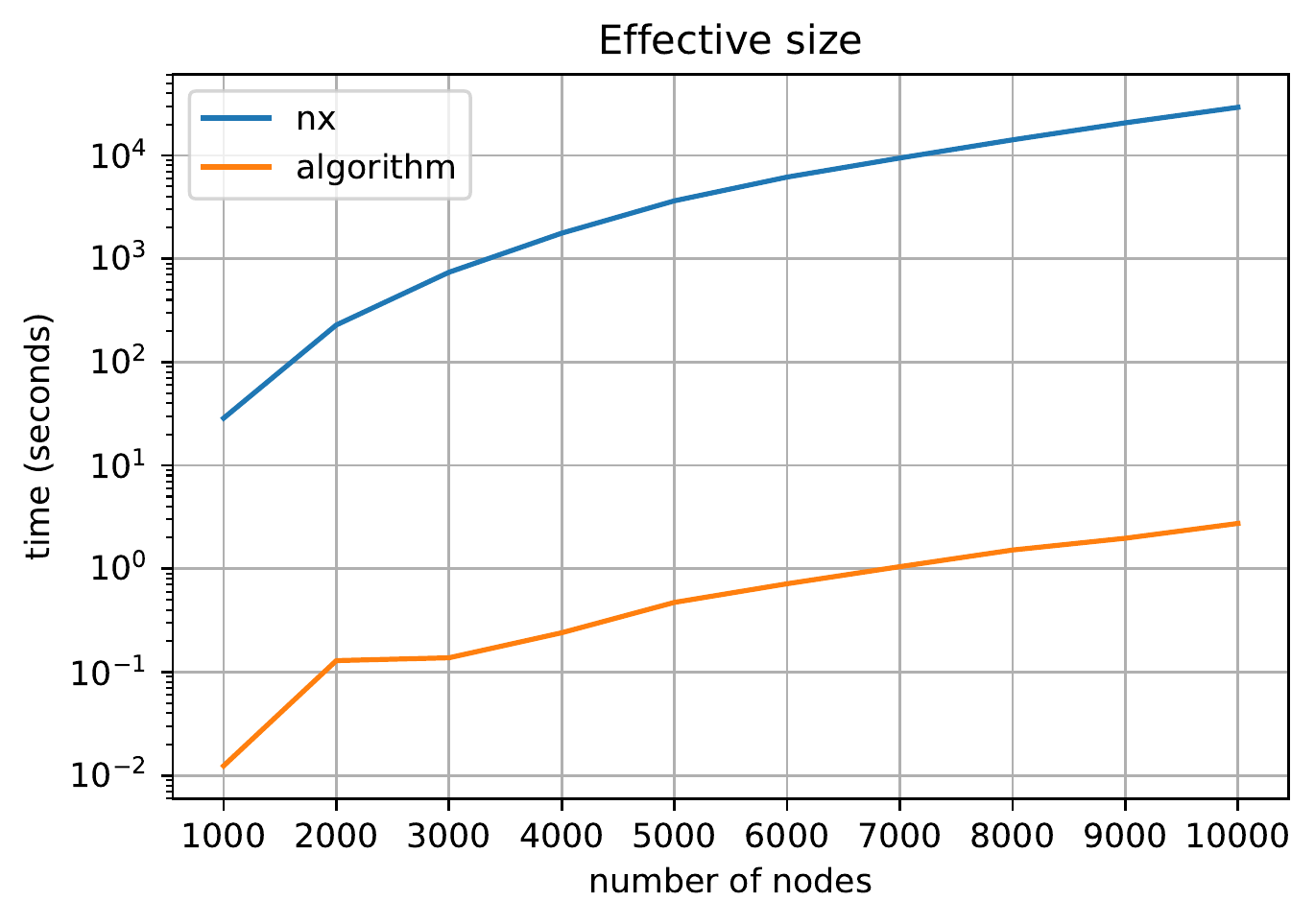}
	\end{subfigure}
	\begin{subfigure}[t]{.49\textwidth}
		\includegraphics[width=\textwidth]{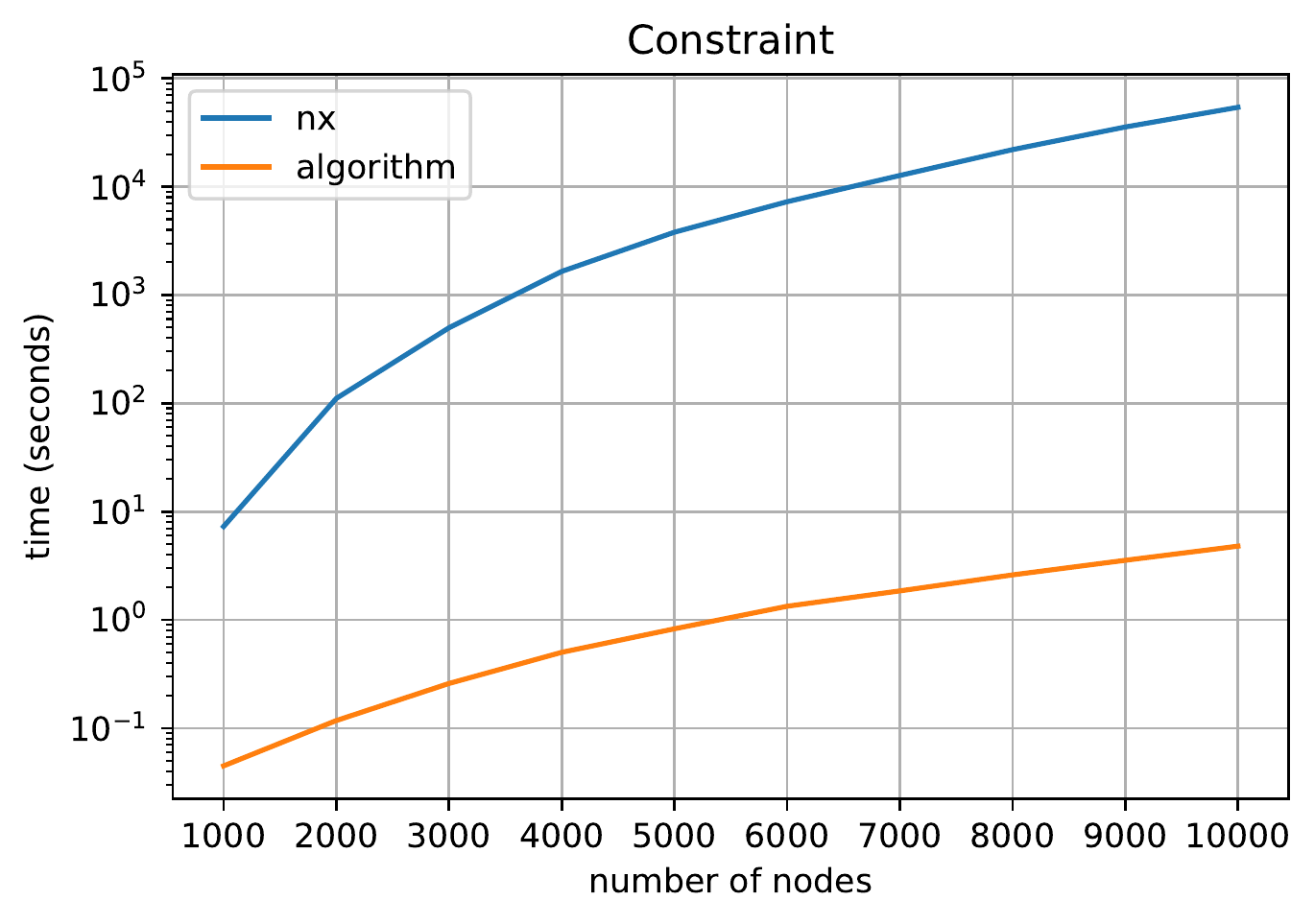}
	\end{subfigure}
	\caption{\small A basic comparison of computational speed between this paper's algorithms (orange) and \href{https://networkx.org/}{NetworkX}'s algorithms (blue) for \href{https://networkx.org/documentation/stable//reference/algorithms/generated/networkx.algorithms.structuralholes.effective\_size.html\#networkx.algorithms.structuralholes.effective\_size}{effective size} (left) and \href{https://networkx.org/documentation/stable//reference/algorithms/generated/networkx.algorithms.structuralholes.constraint.html\#networkx.algorithms.structuralholes.constraint}{constraint} (right), as the number of nodes increases from 1,000 to 10,000. The networks are \href{https://networkx.org/documentation/stable//reference/generated/networkx.generators.random\_graphs.barabasi\_albert\_graph.html}{Barabasi-Albert graphs} with parameter $m=5$ (top row) and \href{https://networkx.org/documentation/stable//reference/generated/networkx.generators.random_graphs.erdos\_renyi\_graph.html\#networkx.generators.random\_graphs.erdos\_renyi\_graph}{Erdos-Renyi random graphs} with parameter $p=0.01$ (bottom row). Notice, however, that a precise comparison in a controlled environment should show that more efficient methods scale better and are better suited for large networks.}
	\label{fig:comparison_speed}
\end{figure}

In this note, I consider the main measures associated with structural holes, namely effective size, redundancy, local constraint and constraint, and derive simple formulas from the adjacency matrix of the network.
% This can help to interpret these measures and also produces intuitive and naive algorithms based on matrix multiplications that work fairly well on moderately large networks (see Figure \ref{fig:comparison_speed}), without the need to rely on more advanced techniques for triangle listing in sparse graphs \citep{chiba1985arboricity} and distributed algorithms.
This might help to interpret these measures and also produces intuitive and naive algorithms based on matrix multiplications which work fairly well on moderately large networks (see Figure \ref{fig:comparison_speed}). 
However, while this approach is clean and simple, it has clear limitations when the network under analysis is very large. 
In such a case, one should avoid storing explicitly the matrices and should preferably rely on distributed algorithms and more advanced techniques for triangle listing with vertex orderings and neighborhood markers \citep{chiba1985arboricity,li2019distributed}.%\footnote{One of the limiting factors of this paper's simple approach is that the matrix $A^2$ is denser than the adjacency matrix $A$. A comparison in a controlled environment should show that more efficient algorithms based on triangle listing with vertex orderings and neighborhood markers scale better with the number of nodes than the naive algorithms proposed here.}

\section{Notation}

In what follows, matrices are denoted by capital letters (e.g. $A$, $P$) and their elements denoted by the corresponding letter with subscripts (e.g. $a_{ij}$, $p_{ij}$). 
Generic nodes of a network (i.e., a graph) will be indicated by $i$, $j$ or $k$. Consequently, an adjacency matrix will be indicated by $A = (a_{ij})_{i,j=1,...,n}$, where $n$ is the number of nodes and the elements $a_{ij}$ can be 0 or 1 for binary networks or generic real numbers for weighted networks.

Vectors and their elements will be respectively denoted by bold letters (e.g. $\mathbf x$, $\mathbf y$) and letters with a (single) subscript (e.g. $x_i$).
The vector obtained by taking the diagonal elements of a square matrix $A$ is denoted by $Diag(A)$ and, analogously, the matrix that has $\mathbf x$ as its diagonal and 0s elsewhere is denoted by $Diag(\mathbf x)$.
The transposed of a vector or matrix is denoted by $\cdot^T$ (e.g. $\mathbf x^T$, $A^T$).
Hereafter, vectors are considered as columns, that is ($n \times 1$)-matrices and their transposed as row vectors $\mathbf x^T$. 
Accordingly, the (matrix) multiplication of a column vector $\mathbf x$ times a row vector $\mathbf y^T$ will give a matrix (e.g. $\mathbf x \mathbf y^T \in \mathbb R^{n\times n}$) whereas $\mathbf x^T \mathbf y$ a scalar.

The matrix multiplication between two matrices $A$ and $B$ will be denoted by juxtaposition, i.e. $A B$, whereas element-by-element operations such as element-wise multiplication or division will be denoted respectively by $\odot$ and $\oslash$.  
The $n$-dimensional unitary vectors in $\mathbb R^n$ containing all 0s but one 1 in $i$-th position is denoted by $\mathbf e_i$, while the vector containing all 1s is denoted by $\mathbf 1$.
The identity matrix is denoted by $I$.

\section{Effective size and redundancy for undirected binary networks}

\begin{figure}[tb]
    \centering
    \caption*{\bf Description of redundancy and effective size}
    \includegraphics[width=.45\textwidth]{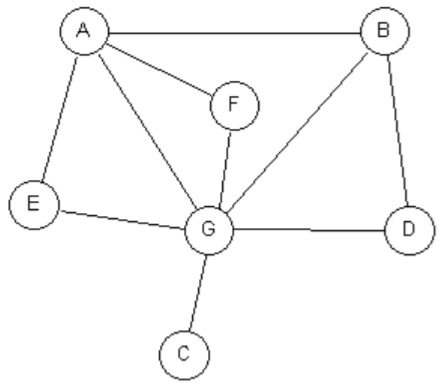}
    \caption{\small Adapted from Burt's and Borgatti's works. Let us compute the effective size for node $A$. Consider one of its neighbors, say $G$. Then, $A$ and $G$ have 3 ``common neighbors'': $B$, $E$ and $F$. Analogously for all 4 neighbors of $A$ (respectively, $B$, $E$, $F$ and $G$). Summing up all these numbers and divide them by $A$'s degree, which is 4, gives: $\frac{1 + 1 + 1 + 3}{4} = \frac{6}{4} = 1.5$. Lastly, $A$'s effective size is its degree minus its redundancy: $4 - 1.5 = 2.5$. In the example of this section, the matricial computation is also done for the remaining nodes.
    % Adapted from Burt's and Borgatti's works. Let us compute the effective size for node $A$. Consider one of its neighbors, say $G$. They have 3 ``common neighbors'': $B$, $E$ and $F$. Divide this number by $A$'s degree: $3/4$. Take another of $A$'s neighbors, say $E$. $A$ and $E$ have just 1 common neighbor, which is $G$. So, in this case dividing by $A$'s degree gives $1/4$. To get $A$'s \emph{redundancy}, repeat this process for all 4 neighbors of $A$ (respectively, $B$, $E$, $F$ and $G$) and sum up all the numbers obtained: $\frac{1}{4} + \frac{1}{4} + \frac{1}{4} + \frac{3}{4} = \frac{6}{4} = 1.5$. Lastly, $A$'s effective size is its degree minus its redundancy: $4 - 1.5 = 2.5$. In the example of this section, the matricial computation is also done for the remaining nodes.
    }
    \label{fig: burt_network}
\end{figure}

The original definition of effective size and redundancy in Burt's works was complicated, but \citet{borgatti1997structural} has shown that it can be simplified. Here, we consider an undirected and binary network with no self-loops. 
The intuitive idea (see Figure \ref{fig: burt_network}) is first to compute a node's \emph{redundancy}, which is the mean number of connections from a neighbor to other neighbors.
Then, the \emph{effective size} is obtained by subtracting the redundancy to the node's degree.

Let $r_i$ be the redundancy of node $i$ and let $t_i$ be ``the number of ties in the network (not including ties to ego)'' \citep{borgatti1997structural}.\footnote{Note: ``ego'' here is node $i$.}
Then, the redundancy is simply\footnote{Since the network is assumed undirected, the links of $t_i$ have to be counted twice.}
\begin{equation}
\label{eq:redundancy}
r_i = \frac{2 \ t_i}{d_i},
\end{equation}
where $d_i$ is $i$'s degree. Notice that $r_i$ goes from 0 to $d_i - 1$.\footnote{It is also well related to the notion of \emph{local clustering}, which can be thought of as a normalized version of redundancy ranging from 0 to 1. It can easily be shown that the relationship between the local clustering $q_i$ and redundancy $r_i$ of a node $i$ of degree $d_i$ is given by: $q_i = \frac{r_i}{d_i - 1}$ \citep{newman2018networks}.}
Then, the effective size $s_i$ of node $i$ is defined as:
\begin{equation}
\label{eq:effective_size}
s_i = d_i - r_i.
\end{equation}

Now, let us see how to compute this in a matricial form.
Let $A = (a_{ij})_{i,j}$ be the adjacency matrix of such an undirected and binary network and let $\mathbf d = (d_i)_i$ be the vector of nodes' degrees.\footnote{$A$ is a symmetric matrix only containing 0s and 1s. In such a case the vector of nodes' degree can be obtained in several ways, for example as $\mathbf d = Diag(A^2)$ or $\mathbf d = A \mathbf 1$.} 
Notice that for a binary network, the elements of the square $A^2$ count the number of common neighbors. Indeed, for every two nodes $i$ and $j$, the $(i,j)$-th element of $A^2$ is:
\begin{equation}
\begin{aligned}
(A^2)_{ij} 
& = \sum_{k = 1}^n a_{ik} a_{kj} = |\{k:\, k \in N(i) \text{ and } k \in N(j)\}| \\
& = \text{number of common neighbors of } i \text{ and } j,
\end{aligned}
\end{equation}
since $a_{ik}$ is different from 0 if and only if $i$ and $k$ are linked and, analogously, $a_{kj}$ is different from 0 if and only if $k$ and $j$ are linked. 
Obviously, we only want to count the common neighbors for pairs of nodes that are actually linked in the network. 
To do so, it suffices to multiply $A^2$ element by element for $A$ itself. Lastly, we want to sum all these numbers and divide them by the corresponding degree.

Summing up, a matricial way to compute the vector of nodes' effective size, $\mathbf s = (s_i)_i$, is by computing the following vector:
\begin{equation}
\label{eq: effective_size}
\mathbf s = 
\mathbf d - (A^2 \odot A) \boldsymbol 1 \oslash \mathbf d,
\end{equation}
where $A^2$ is $A$ squared with the standard matrix multiplication. The $i$-th component of such a vector, $s_i$, is node $i$'s effective size. By definition, the redundancy is just the last term, that is $\mathbf r = (A^2 \odot A) \boldsymbol 1 \oslash \mathbf d$, where $\mathbf r = (r_i)_i$.

% \begin{figure}[tb]
% \caption*{\bf Computational speed}
% \begin{minipage}{0.5\textwidth}
% \includegraphics[width=\textwidth]{speed_effective_size_BA_networks.pdf}
% \end{minipage}
% \hfill
% \begin{minipage}{0.5\textwidth}
% \includegraphics[width=\textwidth]{speed_constraint_BA_networks.pdf}
% \end{minipage}
% \caption{\small Comparison of computational speed between the author's algorithms (orange) and \href{https://networkx.org/}{NetworkX}'s algorithms (blue) for \href{https://networkx.org/documentation/stable//reference/algorithms/generated/networkx.algorithms.structuralholes.effective\_size.html\#networkx.algorithms.structuralholes.effective\_size}{effective size} (left) and \href{https://networkx.org/documentation/stable//reference/algorithms/generated/networkx.algorithms.structuralholes.constraint.html\#networkx.algorithms.structuralholes.constraint}{constraint} (right), as the number of nodes increases from 1,000 to 10,000. The networks are \href{https://networkx.org/documentation/stable//reference/generated/networkx.generators.random\_graphs.barabasi\_albert\_graph.html}{Barabasi-Albert graphs} with parameter $m=5$. The scale on the y-axis (time) is logarithmic, showing that the author's algorithms are several orders of magnitude faster.}
% \label{fig:comparison_speed}
% \end{figure}

\paragraph{Example}
Consider the network in Figure \ref{fig: burt_network}. 
The adjacency matrix and the degree vector are, respectively\footnote{For simplicity, in $A$ the 0s are not indicated. Notice that self loops are not allowed.}
$$
A = 
\begin{pmatrix}
.   &   1   &   .   &   .   &   1   &   1   &   1   \\
1   &   .   &   .   &   1   &   .   &   .   &   1   \\
.   &   .   &   .   &   .   &   .   &   .   &   1   \\
.   &   1   &   .   &   .   &   .   &   .   &   1   \\
1   &   .   &   .   &   .   &   .   &   .   &   1   \\
1   &   .   &   .   &   .   &   .   &   .   &   1   \\
1   &   1   &   1   &   1   &   1   &   1   &   .   \\
\end{pmatrix},
\quad
\mathbf d = 
\begin{pmatrix}
4   \\
3   \\
1   \\
2   \\
2   \\
2   \\
6
\end{pmatrix}.
$$
Now, since 
$$
A^2 = 
\begin{pmatrix}
4   &   1   &   1   &   2   &   1   &   1   &   3   \\
1   &   3   &   1   &   1   &   2   &   2   &   2   \\
1   &   1   &   1   &   1   &   1   &   1   &   0   \\
2   &   1   &   1   &   2   &   1   &   1   &   1   \\
1   &   2   &   1   &   1   &   2   &   2   &   1   \\
1   &   2   &   1   &   1   &   2   &   2   &   1   \\
3   &   2   &   0   &   1   &   1   &   1   &   6   \\
\end{pmatrix},
$$
then equation \eqref{eq: effective_size} gives the effective size for each node:
$$
\text{nodes }
\left\{
\begin{array}{c}
A   \\
B   \\
C   \\
D   \\
E   \\
F   \\
G   \\
\end{array}
\right\}
\quad
\longrightarrow
\quad
\mathbf s = 
\begin{pmatrix}
2.5     \\
1.667  \\
1       \\
1       \\
1       \\
1       \\
4.667
\end{pmatrix} \text{ nodes' effective size}.
$$

\section{Local constraint (a.k.a. dyadic constraint)}

Let $A = (a_{ij})_{i,j}$ be the adjacency matrix of a network (not necessarily binary or unweighted).\footnote{That is, $A$ is not necessarily symmetric and may contains elements different from 0 and 1. The only assumption here is that no self-loop is allowed, that is, $a_{ii} = 0$ for all nodes $i$.}
Following \citet{everett2020unpacking}, the \emph{local constraint} on $i$ with respect to $j$, denoted $\ell_{ij}$, is defined by\footnote{\label{note_alternative}This is also known as \emph{dyadic constraint}. The definition used in \href{https://networkx.org/}{NetworkX}'s algorithm for \href{https://networkx.org/documentation/stable//reference/algorithms/generated/networkx.algorithms.structuralholes.local\_constraint.html\#networkx.algorithms.structuralholes.local\_constraint}{local constraint} is slightly different. The only modification consists in changing $k \in N(i)\setminus\{j\}$ with $k \in N(j)$.
I discuss how to adapt the matricial algorithm in the additional material available at the link in the first acknowledgements note.}
\begin{equation}
\label{eq: def_local_constraint}
\ell_{ij} = \left(p_{ij} + \sum_{k \in N(i)\setminus\{j\}} p_{ik} p_{kj}\right)^2,
\end{equation}
where $N(i)$ is the set of neighbors of $i$ and $p_{ij}$ is the \emph{normalized mutual weight} of the edges joining $i$ and $j$, that is,
\begin{equation}
\label{eq: p_ij}
p_{ij} = \frac{a_{ij} + a_{ji}}{\sum_{k}(a_{ik} + a_{ki})}.
\end{equation}
Notice that, assuming absence of self-loops, every $p_{ii} = 0$, because $a_{ii} = 0$. 
This implies that the second term in definition \eqref{eq: def_local_constraint} can be written as
\begin{equation}
    \sum_{k \in N(i)\setminus\{j\}} p_{ik} p_{kj} = \sum_{k \in N(i)} p_{ik} p_{kj} - p_{ij} \underbrace{p_{jj}}_{=0} = \sum_{k \in N(i)} p_{ik} p_{kj},
\end{equation}
and, hence, $\ell_{ij}$ becomes
\begin{equation}
\label{eq: re_def_local_constraint}
    \ell_{ij} = \left(p_{ij} + \sum_{k \in N(i)} p_{ik} p_{kj}\right)^2.
\end{equation}

Now, let us focus on $p_{ij}$, writing equation \eqref{eq: p_ij} in matricial terms:\footnote{Notice that the denominator here is the multiplication of a row vector times a column vector, which is a number.}
\begin{equation}
p_{ij} = (A + A^T)_{ij} \cdot \frac{1}{\left((A^T + A)\mathbf e_i\right)^T \mathbf 1},
\end{equation}
and let us define vector $\mathbf x = (x_i)_i$, where
\begin{equation}
x_i = (A + A^T) \mathbf e_i^T \mathbf 1 = \sum_{k} (a_{ik} + a_{ki}),
\end{equation}
Thus\footnote{Notice that $A^T + A$ is always symmetric, even if $A$ is not.}
\begin{equation}
\mathbf x = (A + A^T)^T \mathbf 1 = (A^T + A) \mathbf 1
\end{equation}
and we can consider the vector containing all inverted elements:
\begin{equation}
    \mathbf y = \mathbf 1 \oslash \mathbf x =
    \begin{pmatrix}
    1/x_1   \\
    \vdots  \\
    1/x_n.
    \end{pmatrix}
\end{equation}
Then, define the matrix which only consists on the diagonal being equal to $\mathbf y$, that is, $Diag(\mathbf y)$. 
Now, we can finally compute $P = (p_{ij})_{i,j}$ as follows:\footnote{By pre-multiplying a diagonal matrix, we are just multiplying every row $i$ of $(A + A^T)$ for the corresponding element $y_i$ of the diagonal.}
\begin{equation}
    P = Diag(\mathbf y) \ (A + A^T).
\end{equation}

Now, let us focus on $\ell_{ij}$. 
Consider again the second term of the definition's formula as written in equation \eqref{eq: re_def_local_constraint}
\begin{equation}
\label{eq: local_constraint_summation}
    \sum_{k \in N(i)} p_{ik} p_{kj} = \sum_{k} a_{ik} p_{ik} p_{kj},
\end{equation}
where the summation on the right-hand side is over all nodes $k$ (not just limited to $i$'s neighbors).\footnote{\label{note_weighted_matrix}If the network is weighted, then here one has to first compute the binary version $A$ of the weighted adjacency matrix $W$, where $a_{ij} = 1$ if and only if $w_{ij} \neq 0$ and $a_{ij} = 0$ otherwise. Then, one can just apply the formula written in the text.}
Written in matricial form, this summation in equation \eqref{eq: local_constraint_summation} can simply be expressed as\footnote{The modification mentioned in Footnote \ref{note_alternative} consists in taking here $P(P\odot A)$.}
\begin{equation}
    (A \odot P) P,
\end{equation}
where $\odot$ is the element-wise matricial multiplication and the second is a matrix multiplication.

To conclude, we can write the matrix $L = (\ell_{ij})_{i,j}$ containing all links' local constraints as follows:
\begin{equation}
\label{eq: local_constraint}
    L = \big[P + (A \odot P) P\big] \odot \big[P + (A \odot P) P\big].
\end{equation}
Summing up, the algorithm takes the adjacency matrix $A$ as input and proceeds with the following steps:
\begin{enumerate}
    \item   $\mathbf x = (A + A^T) \mathbf 1$;
    \item   $\mathbf y = \mathbf 1 \oslash \mathbf x$;
    \item   $P = Diag(\mathbf y) \ (A + A^T)$;
    \item   $L = \big[P + (A \odot P) P\big] \odot \big[P + (A \odot P) P\big]$.
\end{enumerate}

\section{Constraint}

% Let $L = (\ell_{ij})_{i,j}$ be the local constraint matrix computed in equation \eqref{eq: local_constraint}. The \emph{constraint on $i$} is defined as:
% \begin{equation}
%     c_i = \sum_{j \in N(i)\setminus\{i\}} \ell_{ij},
% \end{equation}
% where $N(i)$ is the set of $i$'s neighbors.\footnote{In case of a directed network, it can be the set of predecessors or successors of $i$. Notice also that in our notation, $N(i)$ does not include $i$ itself, meaning that the adjacency matrix has all 0s on its diagonal.}

% Then, the vector containing the nodes' constraints, $\mathbf c = (c_i)_i$ can be obtained by summing the rows of $L$ (excluding the elements on its diagonal of the form $l_{ii}$). That is,\footnote{Remember that in our notation vectors are always considered as columns.}
% \begin{equation}
%     \mathbf c = \left(\mathbf 1^T (L - Diag(L)\right)^T.
% \end{equation}

% \hrule

Let $L = (\ell_{ij})_{i,j}$ be the local constraint matrix computed in equation \eqref{eq: local_constraint}.
According to \citet{everett2020unpacking}, the \emph{constraint} for node $i$ is\footnote{Notice that in our notation $N(i)$ does not include $i$ itself. To be even more clear, one could then write $c_i = \sum_{j \in N(i)\setminus\{i\}} \ell_{ij}$.}
$$
c_i = \sum_{j \in N(i)} \ell_{ij}.
$$
One can re-write this as follows:\footnote{In case the network is weighted, then here the matrix $A$ is the binary version of the weighted adjacency matrix $W$, as observed in Footnote \ref{note_weighted_matrix}.}
$$
c_i = \sum_{j} \ell_{ij} a_{ij},
$$
where $(a_{ij})_{i,j} = A$ is the adjacency matrix.
So, the vector $\mathbf c = (c_i)_i$ containing the constraints of the network is obtained by summing the rows of the matrix $L\odot A$:\footnote{Remember that in our notation vectors are always considered as columns.}
$$
\mathbf c = \left[\mathbf 1^T (L \odot A)\right]^T.
$$

\newpage
\bibliographystyle{ecca}
\bibliography{biblio}

\begin{thebibliography}{14}
\providecommand{\natexlab}[1]{#1}

\bibitem[{Bloch \textit{et~al.}(2019)Bloch, Jackson and
  Tebaldi}]{bloch2019centrality}
\textsc{Bloch, F.}, \textsc{Jackson, M.~O.} and \textsc{Tebaldi, P.} (2019).
  Centrality measures in networks. \textit{Available at SSRN 2749124}.

\bibitem[{Borgatti(1997)}]{borgatti1997structural}
\textsc{Borgatti, S.~P.} (1997). Structural holes: Unpacking {B}urt’s
  redundancy measures. \textit{Connections}, \textbf{20}~(1), 35--38.

\bibitem[{Burt(2004)}]{burt2004structural}
\textsc{Burt, R.~S.} (2004). Structural holes and good ideas. \textit{American
  Journal of Sociology}, \textbf{110}~(2), 349--399.

\bibitem[{Burt(2009)}]{burt2009structural}
\textsc{---} (2009). \textit{Structural holes: The social structure of
  competition}. Harvard University Press.

\bibitem[{Chiba and Nishizeki(1985)}]{chiba1985arboricity}
\textsc{Chiba, N.} and \textsc{Nishizeki, T.} (1985). Arboricity and subgraph
  listing algorithms. \textit{SIAM Journal on Computing}, \textbf{14}~(1),
  210--223.

\bibitem[{Everett and Borgatti(2020)}]{everett2020unpacking}
\textsc{Everett, M.~G.} and \textsc{Borgatti, S.~P.} (2020). Unpacking
  {B}urt’s constraint measure. \textit{Social Networks}, \textbf{62}, 50--57.

\bibitem[{Goyal(2018)}]{goyal2018heterogeneity}
\textsc{Goyal, S.} (2018). Heterogeneity and networks. In \textit{Handbook of
  Computational Economics}, vol.~4, Elsevier, pp. 687--712.

\bibitem[{Goyal and Vega-Redondo(2007)}]{goyal2007structural}
\textsc{---} and \textsc{Vega-Redondo, F.} (2007). Structural holes in social
  networks. \textit{Journal of Economic Theory}, \textbf{137}~(1), 460--492.

\bibitem[{Granovetter(1985)}]{granovetter1985economic}
\textsc{Granovetter, M.} (1985). Economic action and social structure: The
  problem of embeddedness. \textit{American Journal of Sociology},
  \textbf{91}~(3), 481--510.

\bibitem[{Jackson \textit{et~al.}(2017)Jackson, Rogers and
  Zenou}]{jackson2017economic}
\textsc{Jackson, M.~O.}, \textsc{Rogers, B.~W.} and \textsc{Zenou, Y.} (2017).
  The economic consequences of social-network structure. \textit{Journal of
  Economic Literature}, \textbf{55}~(1), 49--95.

\bibitem[{Li \textit{et~al.}(2019)Li, Zou, Li, Li and Chen}]{li2019distributed}
\textsc{Li, F.}, \textsc{Zou, Z.}, \textsc{Li, J.}, \textsc{Li, Y.} and
  \textsc{Chen, Y.} (2019). Distributed parallel structural hole detection on
  big graphs. In \textit{International Conference on Database Systems for
  Advanced Applications}, Springer, pp. 519--535.

\bibitem[{Newman(2018)}]{newman2018networks}
\textsc{Newman, M.} (2018). \textit{Networks}. Oxford University Press.

\bibitem[{Rub{\'\i}-Barcel{\'o}(2017)}]{rubi2017structural}
\textsc{Rub{\'\i}-Barcel{\'o}, A.} (2017). Structural holes in social networks
  with exogenous cliques. \textit{Games}, \textbf{8}~(3), 32.

\bibitem[{Schweitzer \textit{et~al.}(2009)Schweitzer, Fagiolo, Sornette,
  Vega-Redondo, Vespignani and White}]{schweitzer2009economic}
\textsc{Schweitzer, F.}, \textsc{Fagiolo, G.}, \textsc{Sornette, D.},
  \textsc{Vega-Redondo, F.}, \textsc{Vespignani, A.} and \textsc{White, D.~R.}
  (2009). Economic networks: The new challenges. \textit{Science},
  \textbf{325}~(5939), 422--425.

\end{thebibliography}

\end{document}